\documentclass[%
preprint,
 amsmath,amssymb,
 aps,
floatfix,
]{revtex4-1}

\usepackage[utf8]{inputenc}
\usepackage[T1]{fontenc}
\usepackage{color}
\usepackage{graphicx} 

\usepackage{url}
\usepackage{comment}

\usepackage{setspace}

\usepackage{multirow}
\usepackage{booktabs}
\usepackage{threeparttable}

\usepackage{csquotes}
\usepackage[colorlinks=true, linkcolor=blue, citecolor=blue, urlcolor=blue]{hyperref}

\DeclareUnicodeCharacter{0301}{*************************************}

\def\dv#1#2{{\frac{d{#1}}{d{#2}}}}
\def\pdv#1#2{{\frac{\partial}{\partial{#2}}{#1}}}
\def\ev#1{\left\langle{#1}\right\rangle}
\def\rfrac#1#2{\frac{#1}{#2}}
\def\arccot{\rm{acot}}
\def\ldef{\mathrel{\mathop:}=}

\begin{document}

\title{Dynamical Phase Transitions in Non-equilibrium Networks}

\author{Jiazhen Liu}
\affiliation{%
Department of Physics, University of Miami, Coral Gables, Florida 33142, USA}%
\author{Nathaniel M. Aden}
\affiliation{%
Department of Physics, University of Miami, Coral Gables, Florida 33142, USA}%
\author{Debasish Sarker}
\affiliation{%
Department of Physics, University of Miami, Coral Gables, Florida 33142, USA}%
\author{Chaoming Song}
\email{c.song@miami.edu}
\affiliation{%
Department of Physics, University of Miami, Coral Gables, Florida 33142, USA}%


\begin{abstract}
Dynamical phase transitions (DPTs) characterize critical changes in system behavior occurring at finite times, providing a lens to study nonequilibrium phenomena beyond conventional equilibrium physics. While extensively studied in quantum systems, DPTs have remained largely unexplored in classical settings. Recent experiments on complex systems, from social networks to financial markets, have revealed abrupt dynamical changes analogous to quantum DPTs, motivating the search for a theoretical understanding. Here, we present a minimal model for nonequilibrium networks, demonstrating that nonlinear interactions among network edges naturally give rise to DPTs. Specifically, we show that network degree diverges at a finite critical time, following a universal hyperbolic scaling, consistent with empirical observations. Our analytical results predict that key network properties, including degree distributions and clustering coefficients, exhibit critical scaling as criticality approaches. These findings establish a theoretical foundation for understanding emergent nonequilibrium criticality across diverse complex systems.
\end{abstract}

\maketitle 

Dynamical phase transitions (DPTs) are critical phenomena that unveil critical behavior emerging over time and represent a new frontier of nonequilibrium physics\cite{heyl2018dynamical,eckstein2009thermalization,garrahan2010thermodynamics,diehl2010dynamical,sciolla2011dynamical,sciolla2013quantum,maraga2016linear,zhang2017observation,heyl2013dynamical}. Distinct from conventional phase transitions, typically driven by external parameters such as temperature or pressure, DPTs occur at a critical time $t_c$, where physical observables display nonanalytic behavior. While these transitions have been extensively studied in quantum systems\cite{maraga2016linear,zhang2017observation,heyl2013dynamical}, their role in classical contexts remains largely underexplored.

Intriguingly, recent experiments on macroscopic complex systems, including ecological collapse\cite{xu2023non,fan2021synchronization}, financial crises\cite{heiberger2014stock,gao2014quantifying}, and power grid failures\cite{fairley2004unruly}, report abrupt dynamical changes at finite times. These events, though not formally classified as DPTs, share key characteristics such as abrupt, collective changes. The striking similarity between these phenomena and quantum DPTs suggests that critical temporal dynamics may be a widespread feature of complex systems. This raises intriguing questions about their broader relevance and underlying universality.

More recently, studies on social networks have presented compelling evidence for genuine DPTs in classical settings\cite{johnson2016new,johnson2019hidden,manrique2018generalized,johnson2020online}. These experiments reveal some physical quantities $z(t)$ diverges at a critical time $t_c$ according to a universal hyperbolic scaling:
\begin{equation}\label{powerlaw} 
z(t) \sim (t_c - t)^{-1}.
\end{equation}
Unlike quantum DPTs, where nonanalyticity emerges without divergence, this classical hyperbolic scaling reveals a distinct signature of temporal criticality. Moreover, this hyperbolic divergence appears to be ubiquitous across distinct domains. For instance, hyperbolic discounting observed in economics\cite{dasgupta2005uncertainty,laibson1997golden,karp2005global,diamond2003quasi}, which rationalizes procrastination and deadlines, also adheres to this scaling, as evidenced by surges in academic submissions near deadlines\cite{durakiewicz2016universal}. Despite its prevalence, the theoretical underpinnings of this universality remain poorly understood.

Existing theoretical frameworks for nonequilibrium classical phase transitions generally fit into two broad categories. The first addresses different phases emerging under time-varied control parameters, much like conventional equilibrium transitions\cite{riego2017metamagnetic,deger2023persistent,canovi2014first,zhang2017observation,diehl2010dynamical,klinder2015dynamical,muniz2020exploring,liang2024thermodynamic,radicchi2020classes,moran2024timeliness}. Some studies even allow these control parameters to vary over time, promoting the system through distinct dynamical regimes\cite{krapivsky2010kinetic,manrique2018generalized}. The second encompasses models of self-organized criticality\cite{bak2013nature,bak1989self,bak1988self,vidiella2021engineering,helmrich2020signatures,wang2013self,garlaschelli2007self}, where critical points occur in the asymptotic limit $t \to \infty$ rather than at a finite $t_c$. Neither category fully captures the finite-time singularities characteristic of DPTs, indicating a need for new theoretical approaches. 

In this study, we introduce a simple theoretical framework that predicts the emergence of DPTs in nonequilibrium networks. Our central insight is that nonlinear interactions play a pivotal role. Drawing an analogy to many-body physics, a random graph without interactions behaves like a system of noninteracting particles with no DPTs. Once nonlinear interactions are introduced, however, the system's collective dynamics are fundamentally altered. To demonstrate this, we consider a simple model involving quadratic interactions, motivated by triangle closure, a ubiquitous mechanism in social networks. Within this model, quadratic interactions between edges naturally yield DPTs. Remarkably, the model can be solved analytically, reproducing the universal scaling behavior expressed in Eq.~\eqref{powerlaw}. Furthermore, we derive critical network characteristics, such as the degree distribution and clustering coefficient, exhibit critical scaling behavior as $t_c$ is approached. These results establish a theoretical basis for understanding nonequilibrium criticality in classical systems, shedding light on parallels with quantum DPTs.

\begin{figure}[!htb]
\centering
    \centering
    \includegraphics[width=1\linewidth]{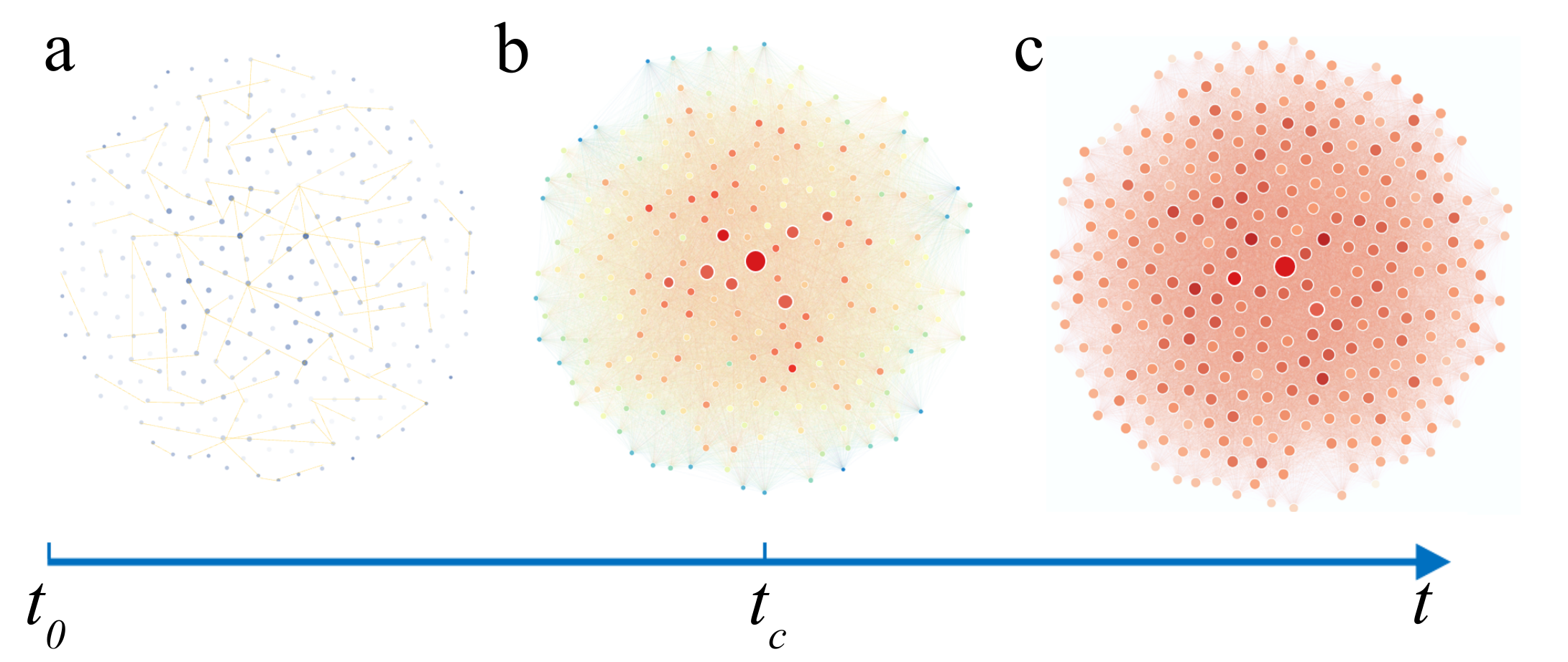}
    \caption{Illustration of network evolution over time for $\beta=0.5$, $\alpha=1$, and $\gamma=1$ with $N=300$: (\textbf{a}) Before the critical time $t_c$, the network exhibits a sparse structure. (\textbf{b}) At $t \to t_c$, the network transitions to a scale-free topology. (\textbf{c}) After $t_c$, the network becomes dense, resembling a nearly complete graph.}
    \label{fig:net}
\end{figure}	

\section*{Theorical framework}
We consider a directed network comprising \(N\) nodes undergoing random edge addition and removal. Specifically, a node \(i\) connects to a node \(j\) to form a new edge with a probability rate \(\alpha/N\), while edges are removed at a rate \(\gamma\). The parameters \(\alpha\) and \(\gamma\) characterize the rates of edge formation and removal, respectively, and the factor \(1/N\) ensures the sparsity of the network. The evolution of the expected average degree, \(z(t)\), follows the linear differential equation:
\[
\dv{z}{t} = \alpha - \gamma z,
\]
with the stationary solution given by \(z(t \to \infty) = \alpha / \gamma\). In this non-interacting scenario, the network remains a random graph, exhibiting no DPTs.

To induce a DPT, it is necessary to introduce interactions. A natural approach is to modify the edge addition rate to \(\alpha/N + U_{int}\), where \(U_{int}(A)\) represents an interaction term dependent on the Boolean adjacency matrix \(A\). While the edge removal rate could also be modified similarly, we hold this parameter constant for simplicity. The interaction term can be expanded as \(U_{int}(A) = \beta_0 A + \beta A^2 + O(A^3)\). To isolate the nonlinear effects, we neglect the linear term and retain only the leading nonlinear contribution, yielding \(U_{int}(A) \approx \beta A^2\), where \(\beta\) is the coupling constant. These interactions are naturally interpreted as a triadic closure mechanism: two nodes, \(i\) and \(j\), sharing a common neighbor are more likely to form a direct connection. This occurs with an additional rate proportional to \(\sum_k A_{ik}A_{kj} = (A^2)_{ij}\). While triadic closure is typically associated with social networks, our theory treats it as the leading nonlinear interaction, extending its applicability beyond social contexts.

\begin{figure}[!htb]
\centering
    \centering
    \includegraphics[width=1\linewidth]{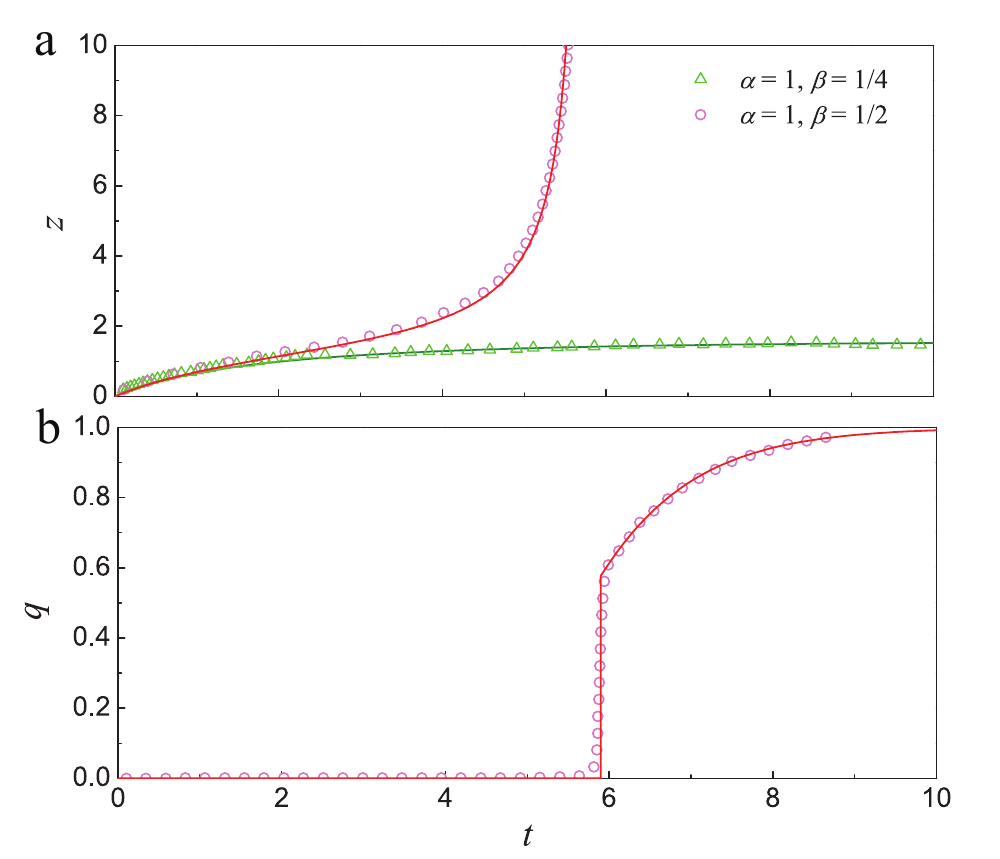}
    \caption{Theoretical predictions versus numerical simulations for (\textbf{a}) the average degree $z(t)$ and (\textbf{b}) the order parameter $q(t)$, with parameters $\alpha = 1$, $\gamma = 1$, $\beta = 1/4$ (green), and $\beta = 1/2$ (red). Solid curves represent theoretical predictions, while scatter points correspond to simulations for $N = 2,000$. For $\beta = 1/4$ (green), the average degree $z(t)$ converges, while the order parameter $q(t)$  vanishes, indicating an equilibrium phase. For $\beta = 1/2$ (red), $z(t)$ exhibits hyperbolic scaling and diverges at a finite critical time $t_c \approx 5.9$. Correspondingly, $q(t)$ shows a sharp jump at $t_c$, signifying a first-oder DPT from a sparse to a dense network.}
\label{fig:zevo}
\end{figure}	

In the weak-interaction regime (\(\beta\) small), the network behaves as a random graph with no DPTs. However, as \(\beta\) increases, nonlinear interactions significantly alter the dynamics, leading to edge clustering and finite-time singularities. Figure \ref{fig:net} illustrates the time evolution of the network for the proposed model for $\beta = 0.5$, $\alpha=1$, and $\gamma=1$ with $N=300$ nodes, which suggests that there is a critical time $t_c$ where the network structure changes significantly. To analyze this analytically, we derive the evolution equation for the average degree \(z(t)\) (see Method Section):
\begin{equation}\label{eq:difeqz}
\dv{z}{t} = \alpha - \gamma z + \beta (z^2 - \Delta),
\end{equation}
where \(\Delta \ldef \frac{1}{N}\ev{\sum_{ijk}A_{ij}A_{jk}A_{ik}}\) measures the triangle density in the network. The kinetic term $\alpha -\gamma z$ is the same as the non-interactive random graph model, whereas the interaction term \(\beta (z^2 - \Delta)\) introduces a feedback loop that amplifies clustering, demonstrating that \(z(t)\) is governed by the interplay of kinetic term and nonlinear interactions.

This differential equation is not closed because \(\Delta\) depends on the number of 4-node rectangular shapes. These 4-node rectangulars, in turn, depend on the number of 5-node shapes, and so forth. This creates an infinite hierarchy of differential equations. For instance,
\begin{equation}\label{eq:difeqz1}
    \dv{\Delta}{t} = \beta(z^2 - \Delta) - \gamma (3 \Delta - \square),
\end{equation}
where \(\square\) represents the number of 4-node shapes. However, these higher-order terms involve coefficients proportional to \(\beta^{n-1}\), thus allowing a perturbative treatment. At the lowest (tree) level, since \(\Delta \sim O(\beta^2)\), we can omit it in Eq. \eqref{eq:difeqz}, resulting in:
\begin{equation}
 z(t) \sim
 \begin{cases}
 \text{constant} & (\beta \leq \frac{\gamma^2}{4\alpha},\; t \to \infty), \\
 1/(t - t_c) & (\beta > \frac{\gamma^2}{4\alpha}, \; t \to t_c),
 \end{cases}
\label{eq:zt}
\end{equation}
where \(t_c\) depends on \(\alpha/\gamma\) and \(\beta/\gamma\) (see Methods Section). Equation \eqref{eq:zt} identifies two distinct phases: In the {\it equilibrium phase}, when \(\beta \leq \frac{\gamma^2}{4\alpha}\), \(z(t)\) approaches a constant value as \(t \to \infty\), indicating that the network converges to a stationary state. In the {\it non-equilibrium phase}, when \(\beta > \frac{\gamma^2}{4\alpha}\), the model predicts a DPT where \(z(t)\) diverges as \(t\) approaches \(t_c\), consistent with the empirical scaling law in Eq.~\eqref{powerlaw}.

Adding higher-order terms does not alter the physical picture described above but introduces corrections to the coupling constant. For instance, considering the contribution of \(\Delta\) from Eq. \eqref{eq:difeqz1}, while omitting \(\square\) terms and applying the adiabatic approximation, results in a correction \(\beta \to \beta' \ldef \frac{3\beta\gamma}{3\gamma+\beta} = \beta - \beta^2/(3\gamma) + O(\beta^3)\). This can be interpreted as a dressed coupling constant due to higher-order corrections. Replacing \(\beta\) with the dressed value \(\beta'\) in Eq. \eqref{eq:zt} provides a precise prediction of \(z(t)\), as shown in Fig. \ref{fig:zevo}a, which agrees with numerical simulations for both the equilibrium and non-equilibrium phases.

\begin{figure}[!htb]
\centering
    \includegraphics[width=1\linewidth]{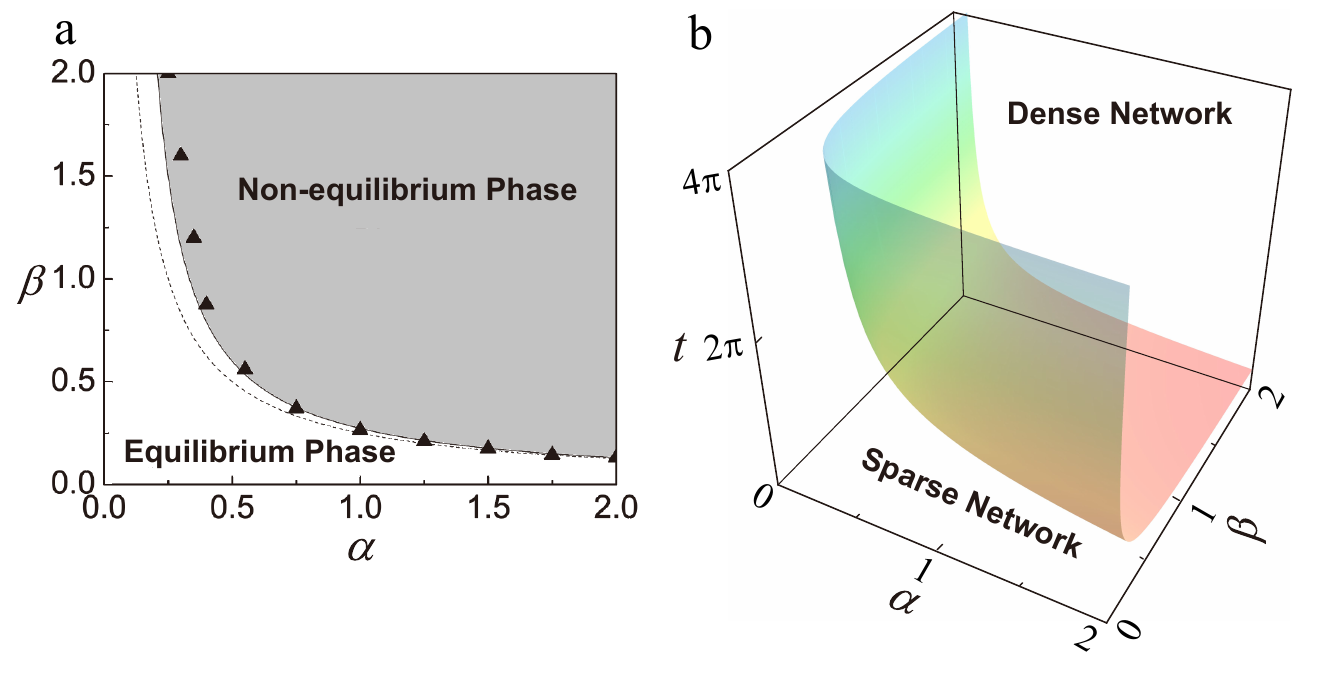}
    \caption{\textbf{(a)} Phase diagram of the model in the $\alpha$-$\beta$ parameter space, showing two distinct phases: equilibrium and non-equilibrium (grey region). The curves represent theoretical predictions, while scatter points correspond to simulation results for the phase boundary. \textbf{(b)} Phase diagram illustrating the transition from sparse to dense graph phases. The colored surface quantifies the phase boundary. Here, $\alpha$, $\beta$, and $t$ are rescaled by $\gamma$ to render them dimensionless.}
\label{fig:blowup}
\end{figure}

Our theoretical analysis above can be summarized in the phase diagram in terms of $\alpha/\gamma$ and $\beta/\gamma$, where the phase boundary $\beta'(\beta)/\gamma = \left(4\alpha/\gamma\right)^{-1}$ separates the equilibrium and non-equilibrium phases. Figure~\ref{fig:blowup}a compares our theoretical prediction with numerical simulations, where the dashed line represents the theoretical prediction at the tree level ($\beta' = \beta$), and the solid line incorporates the triangle correction ($\beta' = \frac{3\beta\gamma}{3\gamma+\beta}$). We find that both predictions agree well with the numerical results for small $\beta$ values, whereas the triangle correction agrees with the numerical results even for relatively large $\beta$.  

Our theory also shows, surprisingly, that small $\beta$ values can still lead to the non-equilibrium phase as long as the kinetic rate $\alpha$ is large enough. This implies that introducing arbitrarily small nonlinear interactions to the random graph model can fundamentally alter the nature of the underlying network evolution, potentially leading to DPT.

For the non-equilibrium phase, the critical time $t_c$ further separates the states into sparse ($t<t_c$) and dense networks ($t>t_c$). The critical time $t_c$ depends on $\alpha$, $\beta$, and $\gamma$, where relatively larger $\beta$ corresponds to a smaller $t_c$, i.e., the system reaches the critical time sooner. For the equilibrium phase, the network will remain sparse for any $t$. Figure~\ref{fig:blowup}b summarizes these facts in the three-dimensional phase diagram. The colored phase boundary surface separates the domains corresponding to sparse and dense networks.

\section*{First-order phase transition}

To further characterize the non-equilibrium phase, we define the order parameter \(q = \lim_{N \to \infty} z/N\), representing the fraction of edges relative to a complete graph. By construction, \(0 \leq q \leq 1\). In the sparse graph regime (\(z \sim O(1)\)), we have \(q = 0\), which holds both in the equilibrium phase and in the non-equilibrium phase before the critical time \(t < t_c\). However, for \(t > t_c\), the network becomes dense, resulting in \(q > 0\). The singularity at \(t = t_c\) in Eq.~\eqref{eq:zt} indicates a first-order phase transition, characterized by a discontinuity, i.e., a latent heat, in \(q(t = t_c)\) from zero to a non-zero value. This behavior is illustrated in Fig.~\ref{fig:zevo}b.

The analytical result in Eq.~\eqref{eq:difeqz1}, however, only captures the regime before the transition (\(t < t_c\)). To determine \(q(t)\) for \(t > t_c\), we need a better understanding of the observed dynamical phase transition (DPT). For sufficiently large coupling, the network is initially sparse, and linear random effects dominate the dynamics. As connections grow, nonlinear interactions arising from triadic closure become increasingly significant, creating competition between linear and nonlinear processes. Once the nonlinear effects dominate, a rapid cascade of edge formation occurs, resulting in a DPT. 

After the DPT, the sparse network condenses into a dense core of a complete graph, accompanied by some scattered isolated nodes of degree zero. Assuming the fraction of isolated nodes is \(p\), the fraction of the complete-graph core is \((1-p(t))N\), and thus \(q(t) = (1-p(t))^2\). As isolated nodes connect to any core member, they rapidly integrate into the core: as soon as an isolated node establishes a single edge to the core, it immediately connects to all other nodes in the core because the number of common neighbors is \(O(N)\), which diverges. Consequently, the rate at which an isolated node connects to a node in the core increases. This implies that \(p(t) = p_0 e^{-\alpha (t-t_c)}\) decays exponentially due to the random connections to the core, where \(p_0\) is the fraction of isolated nodes at \(t = t_c\). From here, we obtain
\begin{equation}
    q(t) = \left(1-p_0 e^{-\alpha (t-t_c)}\right)^2, 
\end{equation}
for \(t \geq t_c\). This prediction agrees well with the numerical simulations (Fig.~\ref{fig:zevo}b). The latent heat \(q(t_c) = (1-p_0)^2\) is related to the critical fraction of isolated nodes \(p_0\), as we will discuss shortly.

\begin{figure}[!htb]
\centering
    \centering
    \includegraphics[width=1\linewidth]{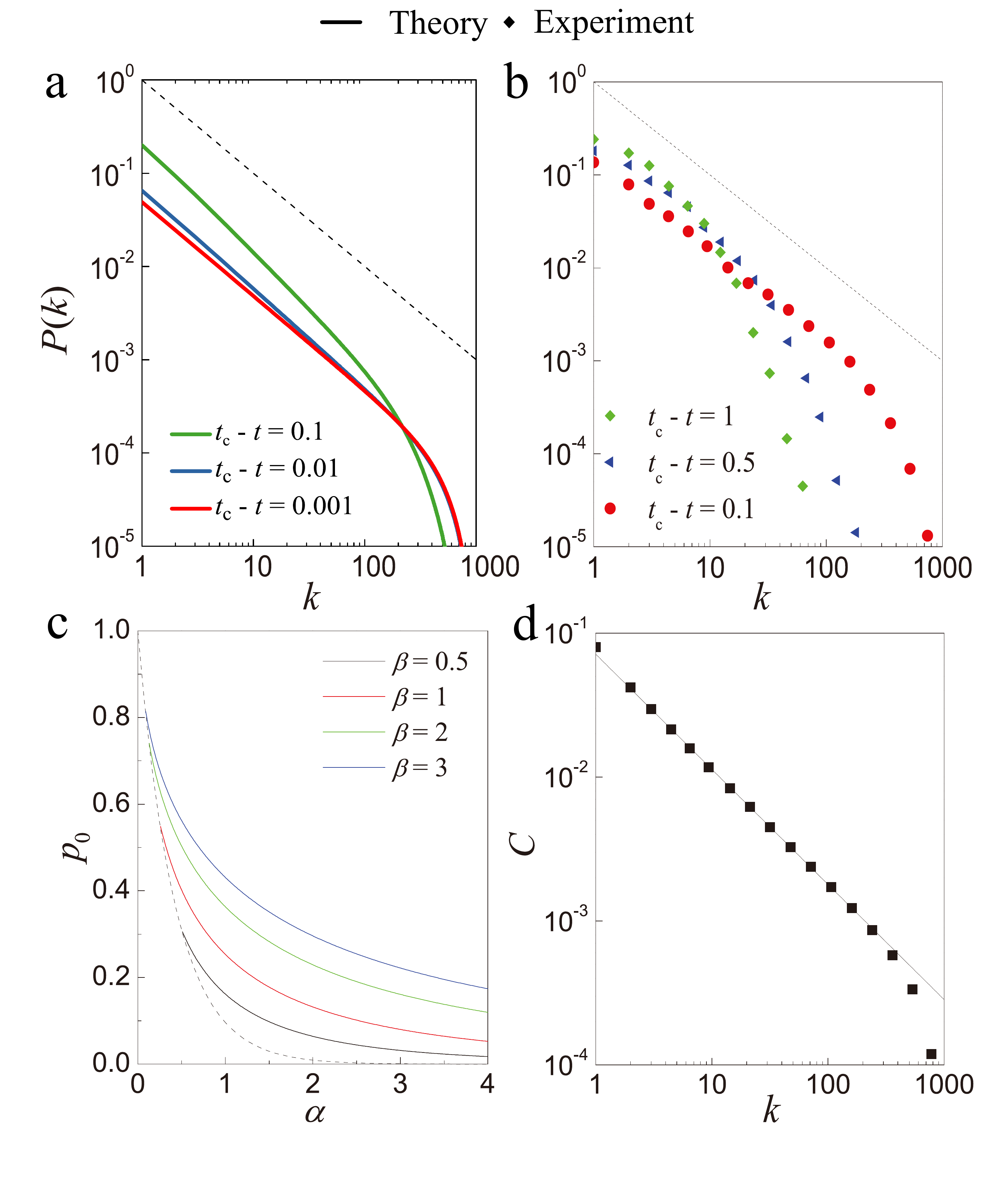}
    \caption{\textbf{Critical Behavior.} (\textbf{a}) Theoretical prediction and (\textbf{b}) numerical simulation of the degree distribution $P(k, t)$ for $t$ near the critical time $t_c$, demonstrating that $P(k)$ follows an asymptotic power law $\propto \frac{1}{k}$ as $t$ approaches criticality. Dashed lines serve as a guide to the eye for the $1/k$ trend. (\textbf{c})Theoretical prediction of the fraction of isolated nodes, $p_0$, at criticality for various parameters. The dashed line indicates the phase boundary between the equilibrium and non-equilibrium phases. (\textbf{d}) The clustering coefficient $C$ as a function of degree $k$ for $t$ near the critical time. Scatter points represent numerical results, while the solid line indicates the theoretical expectation $C(k) \propto \frac{1}{k}$.}    \label{fig:pk}
\end{figure}

\section*{Critical behavior}
We investigate the critical behavior near the DPT. We are particularly interested in the degree distribution. For the non-interactive model (\(\beta = 0\)), the degree distribution is a Poisson distribution. For finite non-linear coupling, at the tree level, we find the time-dependent degree distribution \(P(k,t)\) satisfies 
\begin{align}\label{eq:Pcon1}
&\pdv{P(k,t)}{t}=(\alpha+\beta(k-1)z(t))P(k-1,t)-(\alpha+\beta k z(t)+\gamma k)P(k,t)+\gamma(k+1) P(k+1,t).
\end{align}
This equation can be solved analytically using the generating function method (see Method Section). For the non-equilibrium phase, we find the degree distribution becomes heavy-tailed as it approaches the critical time \(t_c\), as shown in Fig.~\ref{fig:pk}a-b for both theoretical predictions and numerical simulations. Eventually, \(P(k,t\to t_c)\sim k^{-\gamma}\) forms a scale-free network \cite{barabasi1999emergence,albert2002statistical}, with the exponent \(\gamma = 1\), suggesting the divergence of the average degree \(z\), a signal for a dynamical phase transition (DPT). This is particularly interesting because we do not use mechanisms like preferential attachment to generate the scale-freeness \cite{barabasi1999emergence,albert2002statistical}. Instead, the critical behavior near the DPT automatically generates scale-freeness, which lies beyond the conventional paradigm of scale-free networks. 

Moreover, while the tree-level approximation only applies to \(t \leq t_c\), this analytical result allows us to compute the fraction of isolated nodes \(p_0 = P(0, t_c)\). Figure~\ref{fig:pk}c shows the proportion of isolated scattered nodes \(p_0\) under different \(\alpha\) and \(\beta\) while keeping \(\gamma=1\) fixed. The results indicate that stronger nonlinear interactions, i.e., larger \(\beta\), correspond to larger $p_0$, and accordingly smaller latent heat. Thus, it is possible that the first-order DPT could become a continuous phase transition in the strong coupling limit.

The critical behavior is also reflected in other physical quantities. For instance, consider the local clustering coefficient $C_i \ldef \frac{\sum_{j,k}A_{ij}A_{jk}A_{ki}}{k_i(k_i-1)}$ for a node \emph{i} \cite{watts1998collective,allard2024geometric}. Near $t_c$, the nonlinear interaction becomes dominant. The quartic interaction implies that the number of new triangles increases approximately linearly with the number of new edges. Consequently, the number of triangles attached to the node \emph{i} is proportional to its degree, i.e., $\sum_{j,k}A_{ij}A_{jk}A_{ki}\sim k_i$. Equivalently, this leads to $C\sim \frac{1}{k}$ as $t\to t_c$. This theoretical prediction is confirmed by numerical simulations, as shown in Fig.~\ref{fig:pk}d. Such scaling is commonly observed in networks with modular structures, as seen in many real-world networks typically generated by hierarchical mechanisms \cite{ravasz2003hierarchical,ravasz2002hierarchical}. Remarkably, without relying on any of these assumptions, the critical behavior of the DPT naturally results in the same law.

These results demonstrate that, in parallel to self-organized criticality, scaling laws observed in complex systems can also emerge through a DPT. For instance, while scale-free properties are often associated with self-organization, our findings highlight an alternative pathway for such emergent behavior, providing a concrete example of a classical system where DPTs drive these phenomena.

An intriguing question arises from the interplay between classical and quantum DPTs. In quantum systems, the evolution operator \( U = e^{-\int H \, dt} \) is central to understanding phase transitions, linking non-analyticities in the time-dependent rate function to the eigenvalues of \( U \). In classical systems, despite the absence of unitarity, the spectrum of the evolution operator may play an analogous role, potentially leading to DPTs. Our minimal model provides a concrete example of a classical system where DPTs emerge, shedding light on parallels with quantum DPTs and elucidating the underlying mechanisms that may unify these phenomena. The insights gained from this connection have implications far beyond fundamental physics, offering a framework for understanding and addressing critical phenomena in diverse complex systems. 

\section*{Method}
\subsection*{Minimal Model}
For the proposed minimal network model, we consider an unweighted directed network consisting of $N$ nodes, represented by the time-varying adjacency matrix $A(t)$, with matrix elements $A_{ij} = 1$ or $0$, indicating whether node \emph{i} is connected to or disconnected from node \emph{j}. At any time, if there is no edge between nodes $i$ and $j$ ($A_{ij} = 0$), they may establish a new edge with a probability rate $\alpha/N + \beta \sum_k A_{ik}A_{kj}$. Conversely, if $A_{ij} = 1$, i.e., there is an existing edge between $i$ and $j$, it may be removed with a probability rate $\gamma$. The dynamics of this network can be described by the following transition rate matrix:
\begin{subequations}\label{eq:rules}
\begin{equation}\label{subeq:1}
W[A_{ij}(0\to 1)] = \left(\frac{\alpha}{N} + \beta (A^2)_{ij}\right),
\end{equation}
\begin{equation}\label{subeq:2}
W[A_{ij}(1\to 0)] = \gamma, 
\end{equation}
\end{subequations}
where $\alpha$, $\beta$, and $\gamma$ are all non-negative parameters. The average degree $z(t) = \frac{1}{N}\ev{\sum_{ij} A_{ij}}$, where $\ev{\ldots} \ldef \int \ldots P(A,t)dA$. Substituting into Eq.~\eqref{eq:rules} leads to 
\begin{equation}
    \frac{d z}{dt} = \alpha (1-z/N) - \gamma z + \frac{1}{N} \ev{\sum_{ijk} A_{ik} A_{kj}(1-A_{ij})},
\end{equation}
where the term $(1-A_{ij})$ ensures that a new edge will only be established if $A_{ij} = 0$. Since $\ev{\sum_{ij} A_{ik} A_{kj}} = \ev{z_k^{\text{out}}z_k^{\text{in}}}$, and the in- and out-degrees of a single node in our model are uncorrelated, $\frac{1}{N} \langle \sum_{ijk} A_{ik} A_{kj}\rangle = \ev{z^{\text{out}}z^{\text{in}}} = z^2$. Substituting the definition of $\Delta$ and considering only the sparse case $z/N \to 0$, we obtain Eq.~\eqref{eq:difeqz}. 

At the tree level, where we ignore the contributions of $\Delta$, we can find the explicit solution of Eq.~\eqref{eq:difeqz} as
\begin{equation}
\beta z(t) =
 \begin{cases}
 \frac{2\omega}{1-\exp\left[2\sqrt{\beta}\omega t+\ln\left(\frac{\frac{1}{2}+\omega}{\frac{1}{2}-\omega}\right)\right]}+\frac{1}{2}-\omega & (\beta<\frac{\gamma^2}{4\alpha}), \\
 \frac{1}{2}-\frac{\beta}{2\beta+t} & (\beta=\frac{\gamma^2}{4\alpha}), \\
 \frac{\gamma}{2}+\omega \cot\left(\omega(t_c-t)\right) & (\beta>\frac{\gamma^2}{4\alpha}),
\end{cases}
\end{equation}
for the initial condition $z(0) = 0$. Here, $\omega = \sqrt{\lvert\alpha\beta-\rfrac{\gamma^2}{4}\rvert}$ and $t_c=\omega^{-1}\left(\arccot(2\omega)+\pi/2\right)$. A similar result can be obtained for arbitrary initial conditions $z(t=0)$. It is straightforward to verify the asymptotic behavior in Eq.~\eqref{eq:zt} explicitly. 

Finally, we consider the corrections arising from triangles. Omitting \(\square\) terms and applying the adiabatic approximation, Eq.~\eqref{eq:difeqz1} leads to $\beta(z^2 - \Delta) - 3 \gamma \Delta = 0$, or equivalently $\Delta = \frac{\beta}{3\gamma+\beta} z^2$. Substituting this into Eq.~\eqref{eq:difeqz}, we obtain the same equation as the one at the tree level but with a corrected coupling constant $\beta'=\frac{3\beta\gamma}{3\gamma + \beta}$.

\subsection*{Generating Function Method}
To obtain the degree distribution $P(k,t)$, we define the generating function $G(x,t) = \sum_k P(k,t)x^k$. Equation \eqref{eq:Pcon1} leads to the following linear partial differential equation: 
\begin{align}\label{eq:G}
   \pdv{G(x,t)}{t}= (x-1)\left[\alpha G(x,t) - (\gamma-\beta x z(t)) \pdv{G(x,t)}{x}\right].
\end{align}
The general solution of Eq. \eqref{eq:G} is given by
\begin{align}\label{eq:Gsol}
    G(x,t) = &
    \exp \left( \alpha (x-1)\int_0^t \frac{E(t')}{E(t)+(T(t)-T(t'))(x-1)}dt'\right) G_0\left( \frac{x-1}{T(t)(x-1)+E(t)}+1\right), 
\end{align}
where $E(t) \ldef \exp \left( \gamma t -\beta \int_0^t z(t') dt'\right)$, $T(t) \ldef \int_0^t (-\beta z(t')) E(t') dt$, and $G_0(x)$ is the probability generating function at $t = 0$. We assume that the initial graph is empty for simplicity, thus $G_0(x) = 1$. However, similar results can be obtained for general initial conditions. The time-dependent degree distribution $P(k,t)$ can be obtained using the inverse Fourier transformation of Eq. \eqref{eq:Gsol}. To compute the fraction of isolated nodes $p_0$ at criticality, since $p_0 = P(0,t_c) = G(0,t_c)$, together with Eq. \eqref{eq:Gsol}, we obtain   
\begin{align}
    p_0 = 
    \exp \left( -\alpha \int_0^{t_c} \frac{E(t')}{E(t)-(T(t_c)-T(t'))}dt'\right).
\end{align}

\bibliography{ref.bib}

\begin{thebibliography}{10}

\bibitem{heyl2018dynamical}
Markus Heyl.
\newblock Dynamical quantum phase transitions: a review.
\newblock {\em Reports on Progress in Physics}, 81(5):054001, 2018.

\bibitem{eckstein2009thermalization}
Martin Eckstein, Marcus Kollar, and Philipp Werner.
\newblock Thermalization after an interaction quench in the hubbard model.
\newblock {\em Physical review letters}, 103(5):056403, 2009.

\bibitem{garrahan2010thermodynamics}
Juan~P Garrahan and Igor Lesanovsky.
\newblock Thermodynamics of quantum jump trajectories.
\newblock {\em Physical review letters}, 104(16):160601, 2010.

\bibitem{diehl2010dynamical}
Sebastian Diehl, Andrea Tomadin, Andrea Micheli, Rosario Fazio, and Peter
  Zoller.
\newblock Dynamical phase transitions and instabilities in open atomic
  many-body systems.
\newblock {\em Physical review letters}, 105(1):015702, 2010.

\bibitem{sciolla2011dynamical}
Bruno Sciolla and Giulio Biroli.
\newblock Dynamical transitions and quantum quenches in mean-field models.
\newblock {\em Journal of Statistical Mechanics: Theory and Experiment},
  2011(11):P11003, 2011.

\bibitem{sciolla2013quantum}
Bruno Sciolla and Giulio Biroli.
\newblock Quantum quenches, dynamical transitions, and off-equilibrium quantum
  criticality.
\newblock {\em Physical Review B—Condensed Matter and Materials Physics},
  88(20):201110, 2013.

\bibitem{maraga2016linear}
Anna Maraga, Pietro Smacchia, and Alessandro Silva.
\newblock Linear ramps of the mass in the o (n) model: Dynamical transition and
  quantum noise of excitations.
\newblock {\em Physical Review B}, 94(24):245122, 2016.

\bibitem{zhang2017observation}
Jiehang Zhang, Guido Pagano, Paul~W Hess, Antonis Kyprianidis, Patrick Becker,
  Harvey Kaplan, Alexey~V Gorshkov, Z-X Gong, and Christopher Monroe.
\newblock Observation of a many-body dynamical phase transition with a 53-qubit
  quantum simulator.
\newblock {\em Nature}, 551(7682):601--604, 2017.

\bibitem{heyl2013dynamical}
Markus Heyl, Anatoli Polkovnikov, and Stefan Kehrein.
\newblock Dynamical quantum phase transitions in the transverse-field ising
  model.
\newblock {\em Physical review letters}, 110(13):135704, 2013.

\bibitem{xu2023non}
Li~Xu, Denis Patterson, Simon~Asher Levin, and Jin Wang.
\newblock Non-equilibrium early-warning signals for critical transitions in
  ecological systems.
\newblock {\em Proceedings of the National Academy of Sciences},
  120(5):e2218663120, 2023.

\bibitem{fan2021synchronization}
Huawei Fan, Ling-Wei Kong, Xingang Wang, Alan Hastings, and Ying-Cheng Lai.
\newblock Synchronization within synchronization: transients and intermittency
  in ecological networks.
\newblock {\em National science review}, 8(10):nwaa269, 2021.

\bibitem{heiberger2014stock}
Raphael~H Heiberger.
\newblock Stock network stability in times of crisis.
\newblock {\em Physica A: Statistical Mechanics and its Applications},
  393:376--381, 2014.

\bibitem{gao2014quantifying}
Liang Gao, Chaoming Song, Ziyou Gao, Albert-L{\'a}szl{\'o} Barab{\'a}si,
  James~P Bagrow, and Dashun Wang.
\newblock Quantifying information flow during emergencies.
\newblock {\em Scientific reports}, 4(1):3997, 2014.

\bibitem{fairley2004unruly}
Peter Fairley.
\newblock The unruly power grid.
\newblock {\em IEEE Spectrum}, 41(8):22--27, 2004.

\bibitem{johnson2016new}
Neil~F Johnson, Minzhang Zheng, Yulia Vorobyeva, Andrew Gabriel, Hong Qi,
  Nicol{\'a}s Vel{\'a}squez, Pedro Manrique, Daniela Johnson, Eduardo Restrepo,
  Chaoming Song, et~al.
\newblock New online ecology of adversarial aggregates: Isis and beyond.
\newblock {\em Science}, 352(6292):1459--1463, 2016.

\bibitem{johnson2019hidden}
Nicola~F Johnson, Rhys Leahy, N~Johnson Restrepo, Nicholas Vel{\'a}squez,
  Minzhang Zheng, Pedro Manrique, Prajwal Devkota, and Stefan Wuchty.
\newblock Hidden resilience and adaptive dynamics of the global online hate
  ecology.
\newblock {\em Nature}, 573(7773):261--265, 2019.

\bibitem{manrique2018generalized}
Pedro~D Manrique, Minzhang Zheng, Zhenfeng Cao, Elvira~Maria Restrepo, and
  Neil~F Johnson.
\newblock Generalized gelation theory describes onset of online extremist
  support.
\newblock {\em Physical Review Letters}, 121(4):048301, 2018.

\bibitem{johnson2020online}
Neil~F Johnson, Nicolas Vel{\'a}squez, Nicholas~Johnson Restrepo, Rhys Leahy,
  Nicholas Gabriel, Sara El~Oud, Minzhang Zheng, Pedro Manrique, Stefan Wuchty,
  and Yonatan Lupu.
\newblock The online competition between pro-and anti-vaccination views.
\newblock {\em Nature}, 582(7811):230--233, 2020.

\bibitem{dasgupta2005uncertainty}
Partha Dasgupta and Eric Maskin.
\newblock Uncertainty and hyperbolic discounting.
\newblock {\em American Economic Review}, 95(4):1290--1299, 2005.

\bibitem{laibson1997golden}
David Laibson.
\newblock Golden eggs and hyperbolic discounting.
\newblock {\em The Quarterly Journal of Economics}, 112(2):443--478, 1997.

\bibitem{karp2005global}
Larry Karp.
\newblock Global warming and hyperbolic discounting.
\newblock {\em Journal of public economics}, 89(2-3):261--282, 2005.

\bibitem{diamond2003quasi}
Peter Diamond and Botond K{\"o}szegi.
\newblock Quasi-hyperbolic discounting and retirement.
\newblock {\em Journal of Public Economics}, 87(9-10):1839--1872, 2003.

\bibitem{durakiewicz2016universal}
Tomasz Durakiewicz.
\newblock A universal law of procrastination.
\newblock {\em Physics Today}, 69(2):11--12, 2016.

\bibitem{riego2017metamagnetic}
Patricia Riego, Paolo Vavassori, and A~Berger.
\newblock Metamagnetic anomalies near dynamic phase transitions.
\newblock {\em Physical Review Letters}, 118(11):117202, 2017.

\bibitem{deger2023persistent}
Aydin Deger, Aiden Daniel, Zlatko Papi{\'c}, and Jiannis~K Pachos.
\newblock Persistent non-gaussian correlations in out-of-equilibrium rydberg
  atom arrays.
\newblock {\em PRX Quantum}, 4(4):040339, 2023.

\bibitem{canovi2014first}
Elena Canovi, Philipp Werner, and Martin Eckstein.
\newblock First-order dynamical phase transitions.
\newblock {\em Physical Review Letters}, 113(26):265702, 2014.

\bibitem{klinder2015dynamical}
Jens Klinder, Hans Ke{\ss}ler, Matthias Wolke, Ludwig Mathey, and Andreas
  Hemmerich.
\newblock Dynamical phase transition in the open dicke model.
\newblock {\em Proceedings of the National Academy of Sciences},
  112(11):3290--3295, 2015.

\bibitem{muniz2020exploring}
Juan~A Muniz, Diego Barberena, Robert~J Lewis-Swan, Dylan~J Young, Julia~RK
  Cline, Ana~Maria Rey, and James~K Thompson.
\newblock Exploring dynamical phase transitions with cold atoms in an optical
  cavity.
\newblock {\em Nature}, 580(7805):602--607, 2020.

\bibitem{liang2024thermodynamic}
Shiling Liang, Paolo De~Los~Rios, and Daniel~Maria Busiello.
\newblock Thermodynamic bounds on symmetry breaking in linear and catalytic
  biochemical systems.
\newblock {\em Physical Review Letters}, 132(22):228402, 2024.

\bibitem{radicchi2020classes}
Filippo Radicchi, Claudio Castellano, Alessandro Flammini, Miguel~A Mu{\~n}oz,
  and Daniele Notarmuzi.
\newblock Classes of critical avalanche dynamics in complex networks.
\newblock {\em Physical Review Research}, 2(3):033171, 2020.

\bibitem{moran2024timeliness}
Jos{\'e} Moran, Matthijs Romeijnders, Pierre~Le Doussal, Frank~P Pijpers, Utz
  Weitzel, Debabrata Panja, and Jean-Philippe Bouchaud.
\newblock Timeliness criticality in complex systems.
\newblock {\em Nature Physics}, pages 1--7, 2024.

\bibitem{krapivsky2010kinetic}
Pavel~L Krapivsky, Sidney Redner, and Eli Ben-Naim.
\newblock {\em A kinetic view of statistical physics}.
\newblock Cambridge University Press, 2010.

\bibitem{bak2013nature}
Per Bak.
\newblock {\em How nature works: the science of self-organized criticality}.
\newblock Springer Science \& Business Media, 2013.

\bibitem{bak1989self}
Per Bak, Kan Chen, and Michael Creutz.
\newblock Self-organized criticality in the'game of life.
\newblock {\em Nature}, 342(6251):780--782, 1989.

\bibitem{bak1988self}
Per Bak, Chao Tang, and Kurt Wiesenfeld.
\newblock Self-organized criticality.
\newblock {\em Physical review A}, 38(1):364, 1988.

\bibitem{vidiella2021engineering}
Blai Vidiella, Antoni Guillamon, Josep Sardany{\'e}s, Victor Maull, Jordi Pla,
  Nuria Conde, and Ricard Sol{\'e}.
\newblock Engineering self-organized criticality in living cells.
\newblock {\em Nature communications}, 12(1):4415, 2021.

\bibitem{helmrich2020signatures}
S~Helmrich, A~Arias, G~Lochead, TM~Wintermantel, M~Buchhold, S~Diehl, and
  S~Whitlock.
\newblock Signatures of self-organized criticality in an ultracold atomic gas.
\newblock {\em Nature}, 577(7791):481--486, 2020.

\bibitem{wang2013self}
FY~Wang and ZG~Dai.
\newblock Self-organized criticality in x-ray flares of gamma-ray-burst
  afterglows.
\newblock {\em Nature Physics}, 9(8):465--467, 2013.

\bibitem{garlaschelli2007self}
Diego Garlaschelli, Andrea Capocci, and Guido Caldarelli.
\newblock Self-organized network evolution coupled to extremal dynamics.
\newblock {\em Nature Physics}, 3(11):813--817, 2007.

\bibitem{barabasi1999emergence}
Albert-L{\'a}szl{\'o} Barab{\'a}si and R{\'e}ka Albert.
\newblock Emergence of scaling in random networks.
\newblock {\em science}, 286(5439):509--512, 1999.

\bibitem{albert2002statistical}
R{\'e}ka Albert and Albert-L{\'a}szl{\'o} Barab{\'a}si.
\newblock Statistical mechanics of complex networks.
\newblock {\em Reviews of modern physics}, 74(1):47, 2002.

\bibitem{watts1998collective}
Duncan~J Watts and Steven~H Strogatz.
\newblock Collective dynamics of ‘small-world’networks.
\newblock {\em nature}, 393(6684):440--442, 1998.

\bibitem{allard2024geometric}
Antoine Allard, M~{\'A}ngeles Serrano, and Mari{\'a}n Bogu{\~n}{\'a}.
\newblock Geometric description of clustering in directed networks.
\newblock {\em Nature Physics}, 20(1):150--156, 2024.

\bibitem{ravasz2003hierarchical}
Erzs{\'e}bet Ravasz and Albert-L{\'a}szl{\'o} Barab{\'a}si.
\newblock Hierarchical organization in complex networks.
\newblock {\em Physical review E}, 67(2):026112, 2003.

\bibitem{ravasz2002hierarchical}
Erzs{\'e}bet Ravasz, Anna~Lisa Somera, Dale~A Mongru, Zolt{\'a}n~N Oltvai, and
  A-L Barab{\'a}si.
\newblock Hierarchical organization of modularity in metabolic networks.
\newblock {\em science}, 297(5586):1551--1555, 2002.

\end{thebibliography}


\end{document}